\documentclass[12pt,dvips,preprint]{aastex}

\newcommand{\documentname}{\textsl{Article}}
\newcommand{\etal}{\textit{et~al.}}
\newcommand{\project}[1]{\textsl{#1}}

\newcommand{\fig}[1]{Fig.\ \ref{fig:#1}}
\newcommand{\tab}[1]{Table~\ref{tab:#1}}

\newcommand{\eqn}[1]{Eqn.\ \ref{eqn:#1}}

\usepackage{url,amsmath}
\usepackage{algorithm,algorithmic}
\usepackage{graphicx}

\begin{document}

\title{{\em S4}: A Spatial-Spectral model for Speckle Suppression}

\author{Rob~Fergus\altaffilmark{1, 2},
        David~W.~Hogg\altaffilmark{3},
        Rebecca~Oppenheimer\altaffilmark{4},
        Douglas~Brenner\altaffilmark{4},
        Laurent~Pueyo\altaffilmark{5,6}}
\altaffiltext{1}{Department of Computer Science, New York University}
\altaffiltext{2}{to whom correspondence should be addressed: fergus@cs.nyu.edu}
\altaffiltext{3}{Center for Cosmology and Particle Physics, Department of Physics, New York University}
\altaffiltext{4}{Department of Astrophysics, American Museum of Natural History}
\altaffiltext{5}{Space Telescope Science Institute, Baltimore}
\altaffiltext{6}{Dept. of Physics and Astronomy, Johns Hopkins University}

\begin{abstract}
High dynamic-range imagers aim to block out or null light from a very
bright primary star to make it possible to detect and measure far
fainter companions; in real systems a small fraction of the primary
light is scattered, diffracted, and unocculted.  We introduce {\em
  S4}, a flexible data-driven model for the unocculted (and highly speckled) light in
the \project{P1640} spectroscopic coronograph.  The model uses
Principal Components Analysis (PCA) to
capture the spatial structure and wavelength dependence of the
speckles but not the signal produced by any companion. Consequently,
the residual typically includes the companion signal. The
companion can thus be found by filtering this error signal with a
fixed companion model. The approach is sensitive to companions that are of order a percent of the
brightness of the speckles, or up to $10^{-7}$ times the brightness of
the primary star. This outperforms existing methods by a factor of 2-3
and is close to the shot-noise physical limit.
\end{abstract}

\section{Introduction}

High dynamic range imaging and spectroscopy is the next frontier in
the study of exoplanets (\cite{Oppenheimer09,Oppenheimer12,Traub10}).  There
is hope for atmospheric spectroscopy of substantial numbers of giant
planets, direct detection of planets at large albedo and radius, and
eventually---with something like the far-future \project{Terrestrial
  Planet Finder}---direct time-domain imaging and spectroscopy of
Earth-like planets around other stars.  All of these projects involve
exceedingly precise optical designs, in which the light from the
(generally bright) primary star is more-or-less nulled, at least for
certain locations on the focal plane.  Starlight suppression is
achieved using diffraction and carefully designed optics that remove
starlight.  However, this can never be accomplished perfectly because
residual scattered light due to optical path length differences within
the beam cannot be controlled using diffractive techniques.  For
example, simply detecting an earth analog around a star 10 pc away
requires wavefront control on the level of $\lambda/10000$.  Obtaining a
spectrum through the optical and IR would require wavefront control at
the level of $10^{-5} \lambda$, or roughly 10 picometers.  However,
projects are making advances in this direction, with some beginning to
demonstrate nanometer level optical control on real telescopes.  In
any case, these projects and the more ambitious ones planned for the
future require very sophisticated instrument models and software
pipelines that implement them to achieve full sensitivity.

% Perfect nulling is almost never possible, because it requires
% sub-wavelength control of extremely large optical surfaces and the
% wavefronts reflected from them.  There is no future of these projects
% without very sophisticated instrument models and software pipelines
% that implement them.

\begin{figure}[h!]
\begin{center}
\includegraphics[width=6.5in]{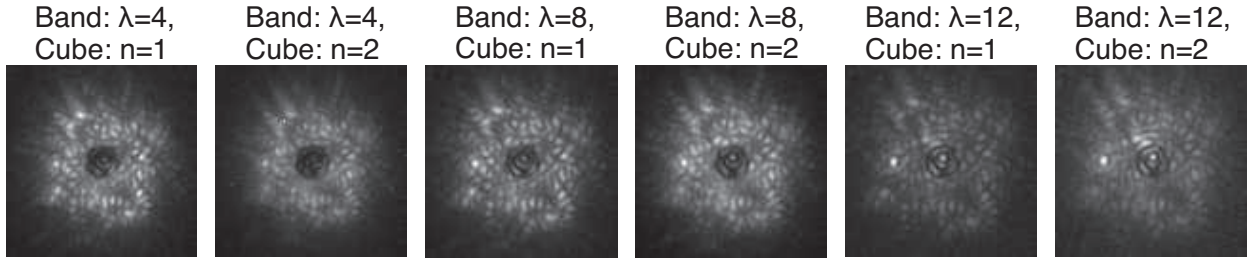}
\end{center}
\vspace{-7mm}
\caption{Examples of 3 different wavelength bands from 2 different
  cubes from the star FU Ori. Note the subtle variation between the
  different cubes, due to changes in atmospheric PSF, as passed
  through the coronograph optical train.}
\label{fig:ex}
\end{figure}

A working example is the \project{P1640} spectroscopic coronograph.
Project 1640 recently demonstrated 5 nm on-sky control of starlight
for observations in the range $\lambda = 0.98 - 1.75 \mu$m
 (\cite{Oppenheimer12}).  This project, the first of several
similar instruments, involves the world's highest order adaptive
optics system (\cite{Dekany06,Roberts12}), an
optimized apodized pupil Lyot coronagraph (\cite{Soummer05}), an
integral field spectrograph (\cite{Hinkley11}), and a calibration
interferometer to detect remnant wave front errors at the starlight
suppression optics (\cite{Wallace10,Zhai12}).  The
project has pioneered many aspects of this complex suite of
instrumentation including data reduction techniques (\cite{Zimmerman11}) and rapid faint companion characterization (\cite{Zimmerman10}).  In addition, the project has explored post-processing
techniques to improve sensitivity (\cite{Crepp11,Pueyo11}).  The images produced by \project{P1640} have the vast
majority of the light from the primary star nulled at the focal plane.
The remaining light falls in a highly speckled pattern produced by
constructive and destructive interference of residual imperfections in
the wavefronts entering the instrument, made worse by differential
chromatic aberration.  The speckle pattern is quasi-random but has
some overall variation with wavelength, somewhat like the pure angular
expansion with wavelength that would be expected for perfect optics,
but not exactly (\cite{Bloemhof01,Racine99,Hinkley07}). See \fig{ex} for two example cubes of FU
ORI at 3 different wavelength bands.

In principle, a wavelength-level model of the wavefronts and all
optical (and non-optical) surfaces in the \project{P1640} system would
produce a precise model of the intensity maps read out at the
detector.  Naively, this model would have of order $10^{12}$
parameters and be intractable to instantiate, let alone fit or
optimize.  Without this model, the speckles are stochastic but with
strong regularities.  This suggests data-driven modeling, or using the
\emph{data that the instrument has in fact taken} to build a precise
and informative but flexible model of the data that it \emph{can}
produce.  If this model can be trained on data that do not have---or
are not expected to have---faint companions contributing, then
companions can be detected as failures of the model, or successes of a
model that is a superposition of the data-driven model and a model
companion.

Of course, we do not know in advance which stars will have companions;
we do not have tagged data for training what are known as
``supervised'' methods.  Relying on the fact that detectable companions are rare, we adopt a ``train and test'' framework, in which we
use, when looking for a companion at a particular location in the
focal plane, all the data in the data set \emph{not} at that location
to train the model.  At the same time, we have very severe precision
requirements, because we are looking for companions that are far
fainter than the residual intensity in the instrument.  This latter
requirement pushes us towards models for the unblocked light that are
extremely flexible, but not so flexible that they can absorb light
from companions.  We will end up getting good performance by using
models with dozens to hundreds of parameters in every small patch of
the focal plane.

In this \documentname, we present a new methodology---{\em S4}---for
modeling high dynamic-range data that is extremely effective when the
images are spectroscopically sliced. The basic principle is to build a
model that jointly captures the spatial and spectral structure of the
data. This is needed to build an effective model of the speckles since
their spatial pattern evolves in a distinctive and predicable way as a
function of wavelength. While a range of models are possible, we adopt
Principal Components Analysis (PCA), also known as the Karhunen-Loeve
transform.  This assumes the data lies on a linear subspace of low
dimensionality, compared to the input dimension. This is a valid assumption for our data as the speckles have
distinctive structure to them, being far from I.I.D. random noise,
hence much of the energy of the signal can be captured by a linear
subspace of a few hundred dimensions at most, far lower than the number of
pixels in the cubes (see \fig{spectrum}(right)). Another advantage,
given the limited number of exposures of a given star, is that PCA
requires relatively little data to build a model.  More complex models
such as sparse coding (\cite{Tibshirani96,Mairal2009}) could also be
used, but requires more data than is typically available in our application.
The full S4 algorithm enables both the detection and spectral analysis
of exoplanets, but this \documentname only address the detection
problem. A forthcoming paper will explain its use for spectral extraction.

{\em S4} benefits from an interdisciplinary collaboration bringing
together astronomy and computer vision.  We demonstrate the method
with \project{P1640} data, and release working code.

\subsection{Related Work}

Several other techniques for speckle suppression have recently been
proposed. LOCI \cite{Lafreniere07} and variants \cite{Pueyo11} attempt to fit the speckles using
a linear combination of patches from nearby wavelengths/exposures.  
As noted by several authors (e.g.~\cite{Marois10},  this approach
risks the possibility of companion flux being accidentally being removed
which impairs performance. Correspondingly, a number of refinements
have been proposed, most notably the SOSIE pipeline \cite{Marois10}. A
notable difference to {\em S4} is that these approaches rely on the
use of Angular Difference Imaging (ADI), which we do not use (since the
telescope on which the P1640 instrument is mounted has an equatorial
mount). In our experiments, we also make a direct comparison to the
damped LOCI algorithm of \cite{Pueyo11}. 

Two recent papers also use PCA to model the speckle
pattern. \project{KLIP} by \cite{Soummer12} has some close similarities
to {\em S4} in that it uses principal components of the observed
data variance to build a reduced-dimensionality description of the
data, which is used in turn as a model.  The principal differences
between \project{KLIP} and {\em S4} arise from the fact that
\project{KLIP} is designed to work on \project{Hubble Space Telescope}
data, which, though variable, do not show as much variation as
ground-based data.  Furthermore, the \project{HST NICMOS} images are
not dispersed, so there is no way to use expected or approximate
variations of speckle patterns with wavelength. Correspondingly,
\project{KLIP} uses PCA to capture the spatial structure of the
speckles, differing from {\em S4} that models the joint {\em
  spatial-spectral} structure. As  
spectral dispersion provides valuable information about the speckle
morphology in each wavelength slice, {\em S4} is better suited to
capturing this structure. 

\cite{Amara12} also propose a PCA based model. In contrast to {\em S4}
which models local image patches with PCA, this work attempts to model
the whole speckle field in with a single PCA basis. Using local model
patches has the advantage that the speckle structure can be captured
with far fewer components that a global model. This work also relies
on the using of ADI.

\section{Method}

While we describe our approach in the context of the \project{P1640}
instrument, the algorithm does not rely on any special properties of
the device, thus can potentially be applied to data produced by other
coronographs of the same type. We do not take advantage of techniques
such as Angular Difference Imaging (ADI), but these could potentially
be combined with our approach to improve performance.

We assume as input to the algorithm properly calibrated intensity information $I_{x,y
, \lambda, n}$ on a four-dimensional boxel grid where the $(x, y)$
indices indicate a pixel number on a regular square pixel grid in the
focal plane, the $\lambda$ index indicates one of a set of narrow
wavelength bins, and the $n$ index is exposure or ``cube'' index 
in the multiple exposures that make up the data set for any one star.
 $I_{x,y, \lambda, n}$ has dimensions $[X \times Y \times \Lambda \times
 N]$. For the \project{P1640} data used in our experiments, $X=Y=101$,
 $\Lambda=32$ and $N \sim 10$, depending on the star.

%The \project{P1640} instrument does not directly deliver data in this
%form; it has to be processed into this form by a non-trivial
%calibration and rectification pipeline (reference) that is outside our
%current scope. 

%  In brief, we are going to build a model for small
% patches of this data by taking representative sets of patches
% (training sets) and building a PCA-based data-driven model for
% patches. In the following explanation we use the star FU Ori as a
% running example, chosen as it has a bright companion embedded in
% speckle artifacts. 

In brief, we are going to build a model for each patch in this data by taking rep-
representative sets of patches (training sets) and building a PCA-based data-driven model for
it. The model is then subtracted from the patch and the residual compared
to test PSF's. In the following explanation we use images of the star
FU Ori as a running example; it has a bright companion embedded in
speckle artifacts that allows easy confirmation of the method's
efficacy. A summary of the approach is given in Algorithm 1.

\subsection{Pre-processing}
The main pre-processing step is the spatial alignment of the data to
ensure that the star is centered within $I_{x,y,\lambda,n}$. Due to
atmospheric dispersion at off-zenith viewing angles, the location of
the star varies with wavelength, thus centering must be performed for
each wavelength $\lambda$, as well as each cube $n$. Currently this is
performed semi-automatically, using the four control spots in the
\project{P1640} data created by diffraction off of a grid of thin
wires in the instrument. The spots are manually identified by
a user for each $\lambda$, $n$ slice of $I_{x,y, \lambda, n}$. The local
maxima around each user click is found and the centroid of the four
maxima computed, which gives the position of the star. Each slice is
then translated (with sub-pixel accuracy) so that the star is
centered.

%
%If guide spots are not present in the data, optical-flow based
%techniques \cite{} can be used to infer the location of the star,
%although this is the subject of future work.

\subsection{Polar transform}

For PCA to be an appropriate model choice, the data must lie on a
linear subspace of low-dimensionality. Correspondingly, we make use of
an important property of the data $I_{x y \lambda n}$, namely that the
evolution of the speckles with wavelength is mostly radial (see Figure
\ref{fig:radiustheta}(bottom) for validation of this). In the
original cartesian representation this radial structure is distinctly
non-linear, thus hard to model with PCA. 
%SOME COMMENT HOW THIS
%COMPARES TO LAURENT AND REMI.

We therefore apply PCA to a polar transformation of the data, instead
of in the original cartesian space. In the polar space, the structure
of the speckles is well captured by a joint 2D model of radius and
$\lambda$, since their tangential motion is small. By contrast, in the cartesian
space, a joint 3D model over $x,y,\lambda$ would be needed. 

\begin{figure}[t!]
\begin{center}
\mbox{
\includegraphics[height=2.8in]{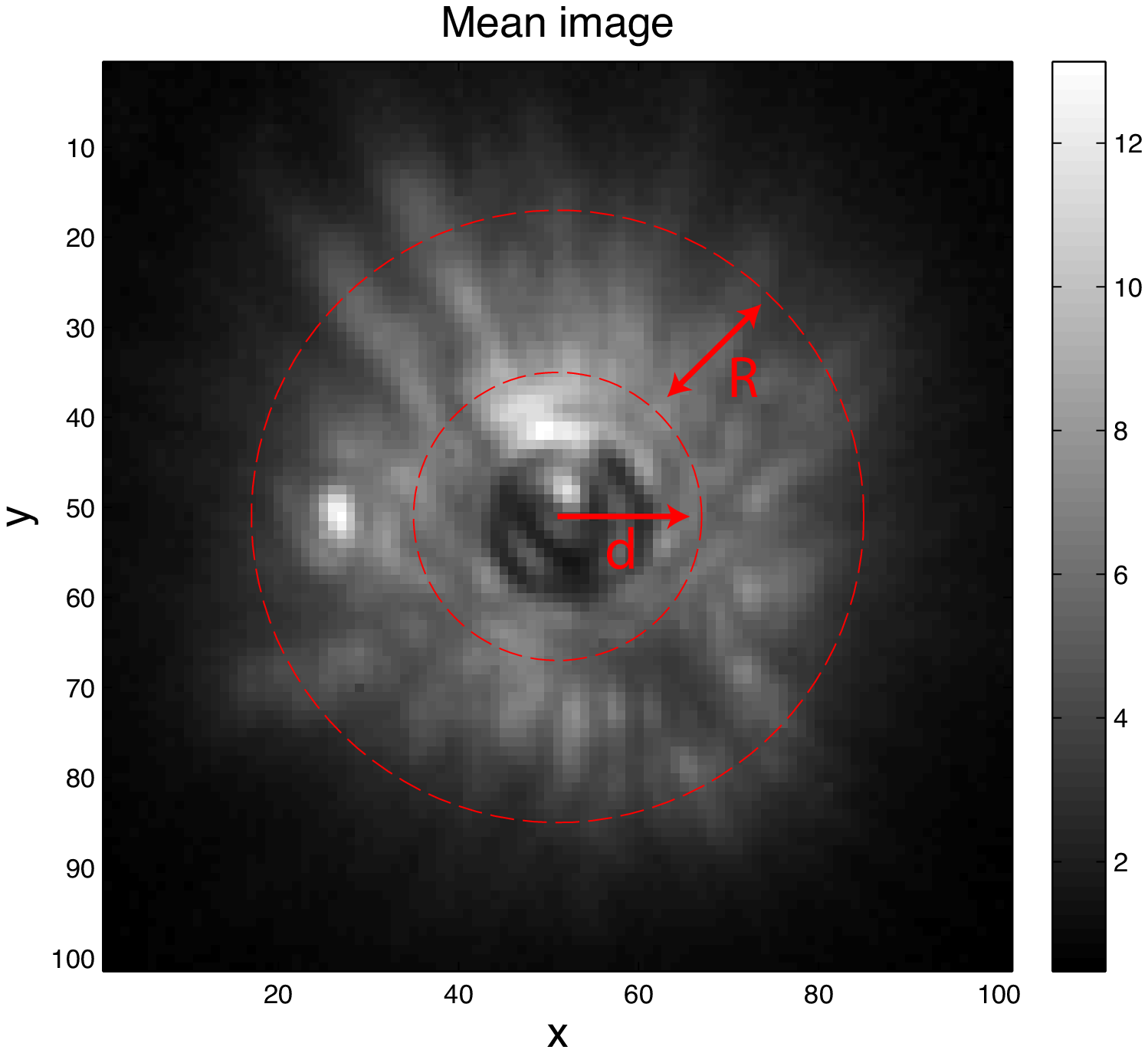}
\hspace{5mm}
\includegraphics[height=2.8in]{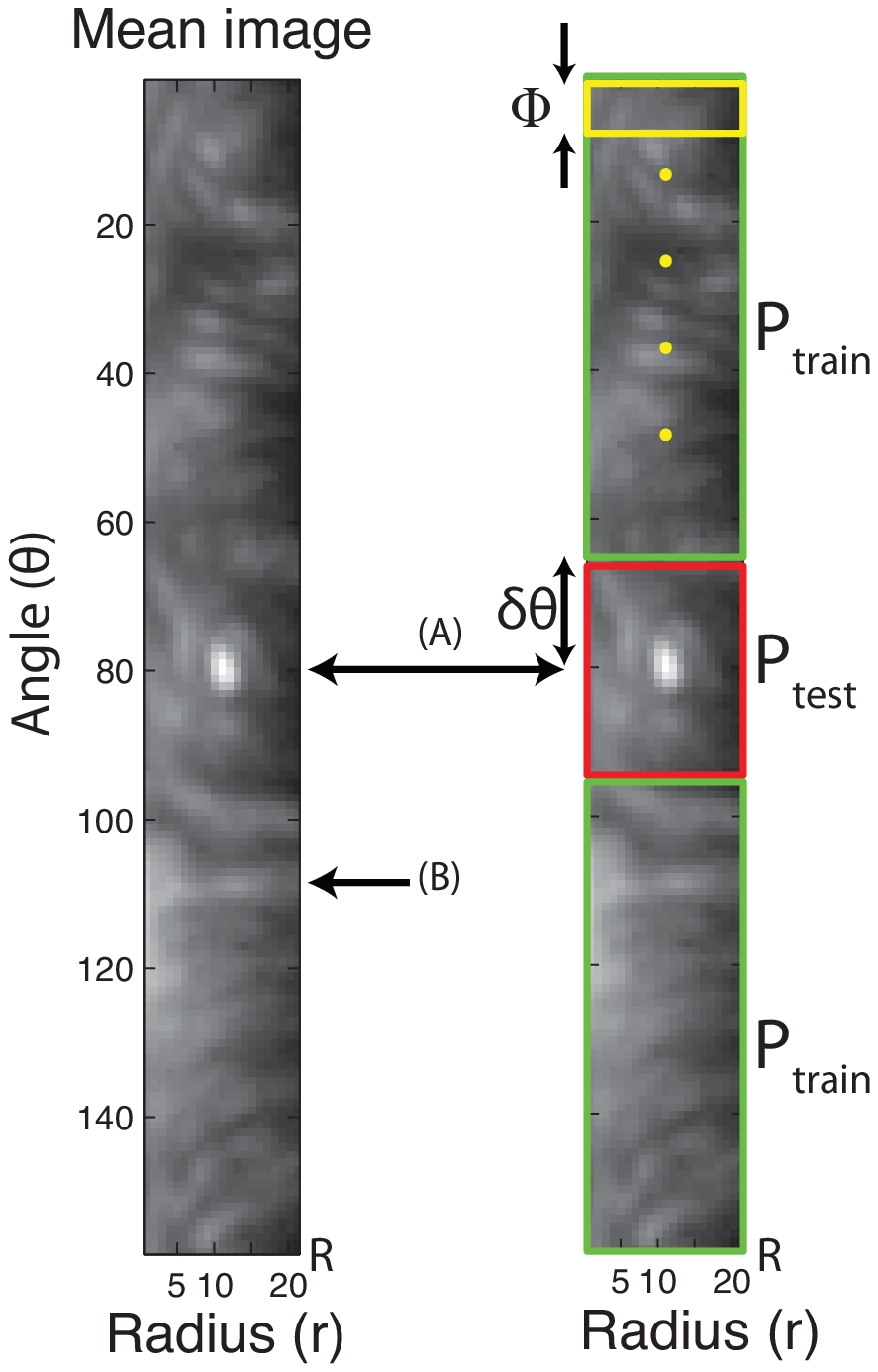}
%\hspace{5mm}
%\includegraphics[height=2.6in]{held_out2}
}
\end{center}
\vspace{-7mm}
\caption{Mean over wavelength $\Lambda$ and cubes $N$ for the star
 FU Ori. {\em Left:} Cartesian representation, with a typical annular
 region highlighted. Note the presence of a
 companion at the 9 o'clock position.  {\em Middle:} Polar representation
 $J$ of the annular region.  {\em Right:} The PCA model is applied to
 overlapping patches (yellow) of angular width $\Phi$ extracted from
 $J$ (also having extent $\Lambda$ over wavelength). We
 test for a companion at a
 given angle, for example, (A) by first training a PCA model on patches
 $P_{train}$ taken outside a zone (of half-width $\delta \theta$)
 around angle (A). The model is then used to reconstruct test patches
 $P_{test}$ near (A). }
\label{fig:mean}
\end{figure}

\begin{figure}[h!]
\begin{center}
\mbox{
\hspace{2mm}
\includegraphics[height=1.8in]{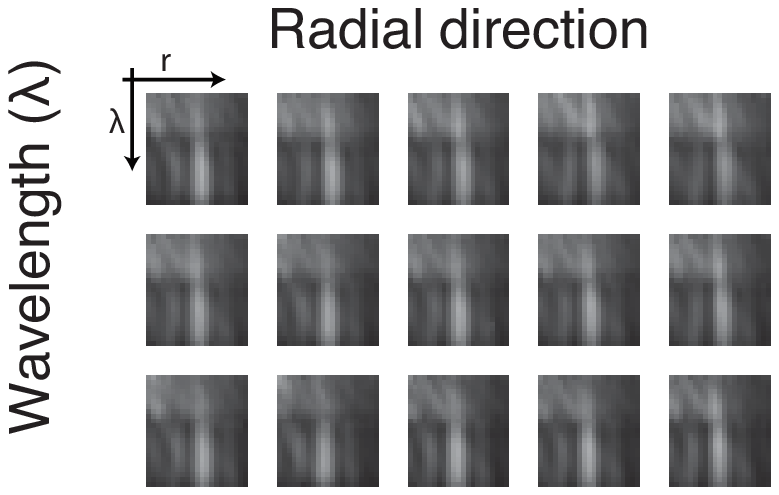}
\hspace{5mm}
\includegraphics[height=1.8in]{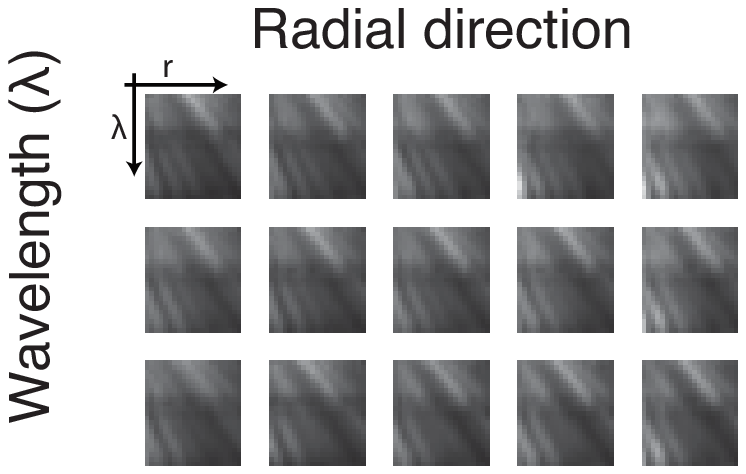}
}\\
\includegraphics[height=2.7in]{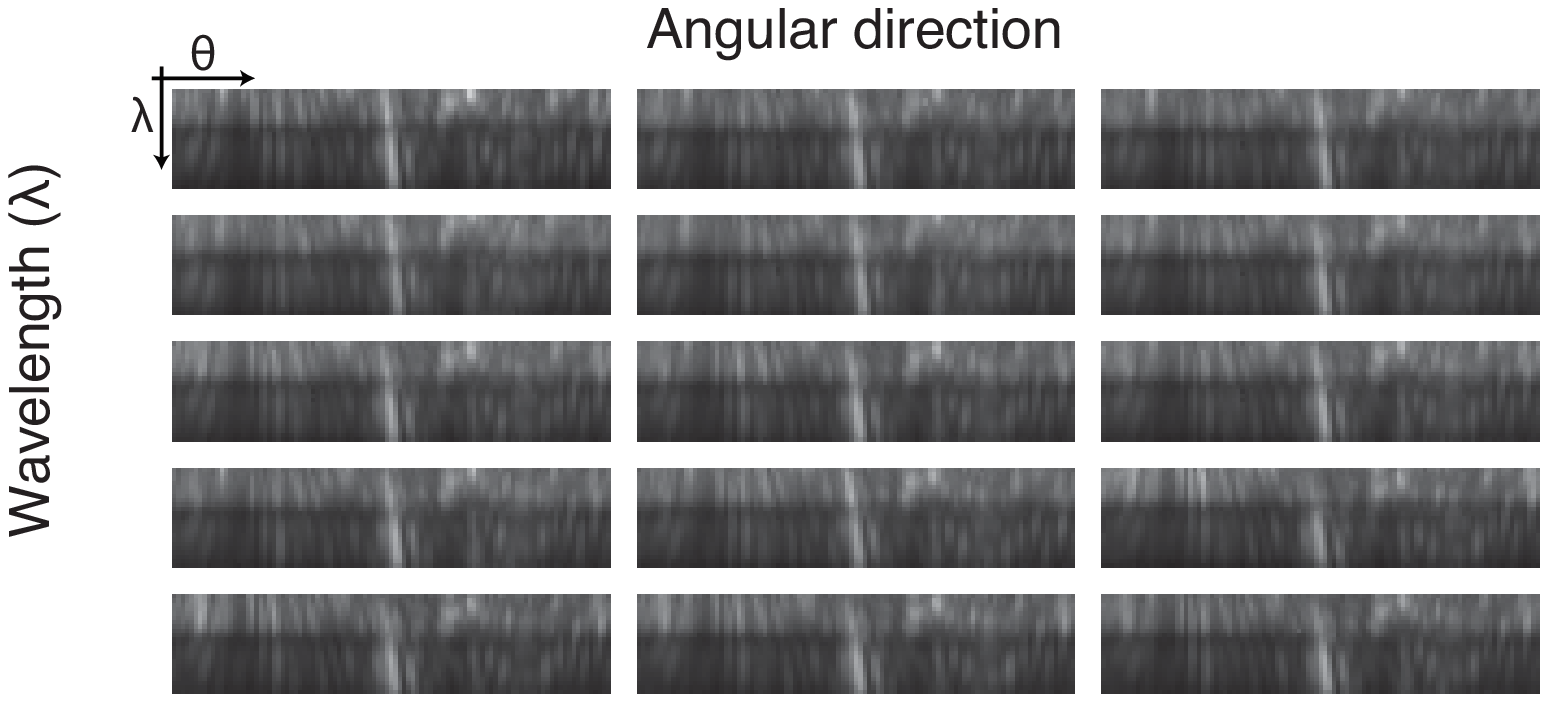}
 
\end{center}
\vspace{-7mm}
\caption{Different visualizations of the polar data volume
  $J_{r,\theta,\lambda,n}$ for the star FU Ori, with $N=15$
  cubes. {\em Top:} Each tile is of dimension $\Lambda \times R$ and shows
  the radial evolution of the speckles in each cube, for two fixed
  angles (left and right), where the companion is and is not present,
  respectively. The angles are indicated by (A) and (B) in
  \fig{mean}(right). The diagonal structure of the speckles contrasts
  with the vertical structure of the companion (whose radius is
  constant with wavelength). In practice the patches $P$ used in our
  model have extent 3 in the angle dimension, rather than 1 as shown
  in these tiles. {\em Bottom:} Each tile is now $\Lambda \times
  \Theta$, for a constant radial location. Note that the angle of the
  speckles is mostly constant as a function of wavelength, i.e.~they
  move radially within the cubes. This justifies our use of patches which are
  small in the $\theta$ dimension. }
\label{fig:radiustheta}
\end{figure}

More precisely, we take centered annular sections of  $I_{x y \lambda n}$,  at
radius $d$, having width $R$ and transform each ($\lambda,n$)
plane separately to a polar representation using bicubic
interpolation, with application of the appropriate Jacobian. The number of
samples $\Theta$ in orientation is $\lceil 2 \pi d\rceil$, chosen to ensure the signal is
well-sampled.  This results in a 4D polar data
array $J_{r,\theta,\lambda,n}$ of dimension $[R \times \Theta \times \Lambda
  \times N]$, where $R \sim 20-30$ typically. \fig{mean} shows
an example of the input data  $I_{x, y, \lambda, n}$  in cartesian representation, along
with the polar projection $J_{r,\theta,\lambda,n}$ of an annular
band, both averaged over $\lambda$ and $n$. \fig{radiustheta} shows
different visualizations of $J_{r,\theta,\lambda,n}$ which reveal the
structure of the speckles.  

\subsection{PCA Patch Model} 

Examining \fig{radiustheta}(bottom), we see that speckle structure
extends over a small angular range, rather than being confined to a
single angle. Thus, to capture the speckle structure, we apply PCA to
{\em patches} extracted from $J_{r,\theta,\lambda,n}$ that have a
narrow window over angle $\theta \pm (\Phi/2)$ and full extent over radius $r$ and
wavelength $\lambda$. For a given angle $\theta$ and cube $n$, each patch $P^{\theta,n}_{r,\phi,\lambda}$ is a 3D volume
of dimension $[R \times \Phi \times \Lambda]$, where the extent over angle
$\Phi$ is small, typically $\sim 3-5$ pix, chosen to (i) reflect the
characteristic diffraction scale and (ii) to keep the patch dimensionality
to a reasonable level. Patches are extracted from
$J_{r,\theta,\lambda,n}$ in ``sliding window'' fashion, i.e.~at every
$\theta$ location, with circular boundary
conditions. $J_{r,\theta,\lambda,n}$ is thus expanded into a set of
$\Theta \cdot N$ patches $P_{r,\phi,\lambda}$, each of dimension $[R \times
\Phi \times \Lambda]$.

At a given radius $d$, we systematically test each angle $\theta$ for
the presence of a companion. This is done by splitting the patches
into disjoint {\em training} and {\em test} sets. The training set
will be used to construct a linear speckle model, based on a PCA eigenvector basis, while the test
set will be fit with those eigenvectors to leave a residual that
potentially may contain a companion. The test set $P_{test}$ is the union of patches
close in angle to $\theta$: $P_{test} = \{P^{\theta,n}\}$ $\forall \, (n,\theta
\pm \delta \theta)$, where $\delta \theta = 5-10$ pixels typically. The
training set $P_{train}$ is the union of patches from all other angles, as
illustrated in \fig{mean}(right).

% The patch resides in a \emph{slice} $S_r$, which is a cutout of size
% $[W\times\Theta\times\Lambda\times N]$ of the data; it consists of the
% unions of all the patches centered on a particular radial index $r$.
% The slice is subdivided into an angular \emph{test region} of angular
% width $\Theta_\test\sim 15$~pix, and the disjoint part of the slice
% makes up the \emph{training region}.  Every patch lies in at least one
% test region (there is a stride $\Delta\Theta$ separating the start
% points of the overlapping test regions in the $\theta$ direction.
% Although the patches have individual angular widths of 3~pix, there is
% a patch defined for every angular bin $1 < \theta > \Theta$.
% Therefore the training set for one patch contains
% $[1\times(\Theta-\Theta_\test)\times 1\times N]$ or about 300 example
% patches.

\begin{figure}[h!]
\begin{center}
\mbox{
\includegraphics[width=3.2in]{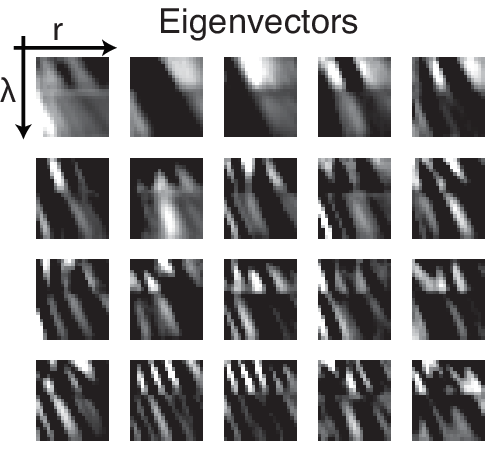}
\includegraphics[width=3.2in]{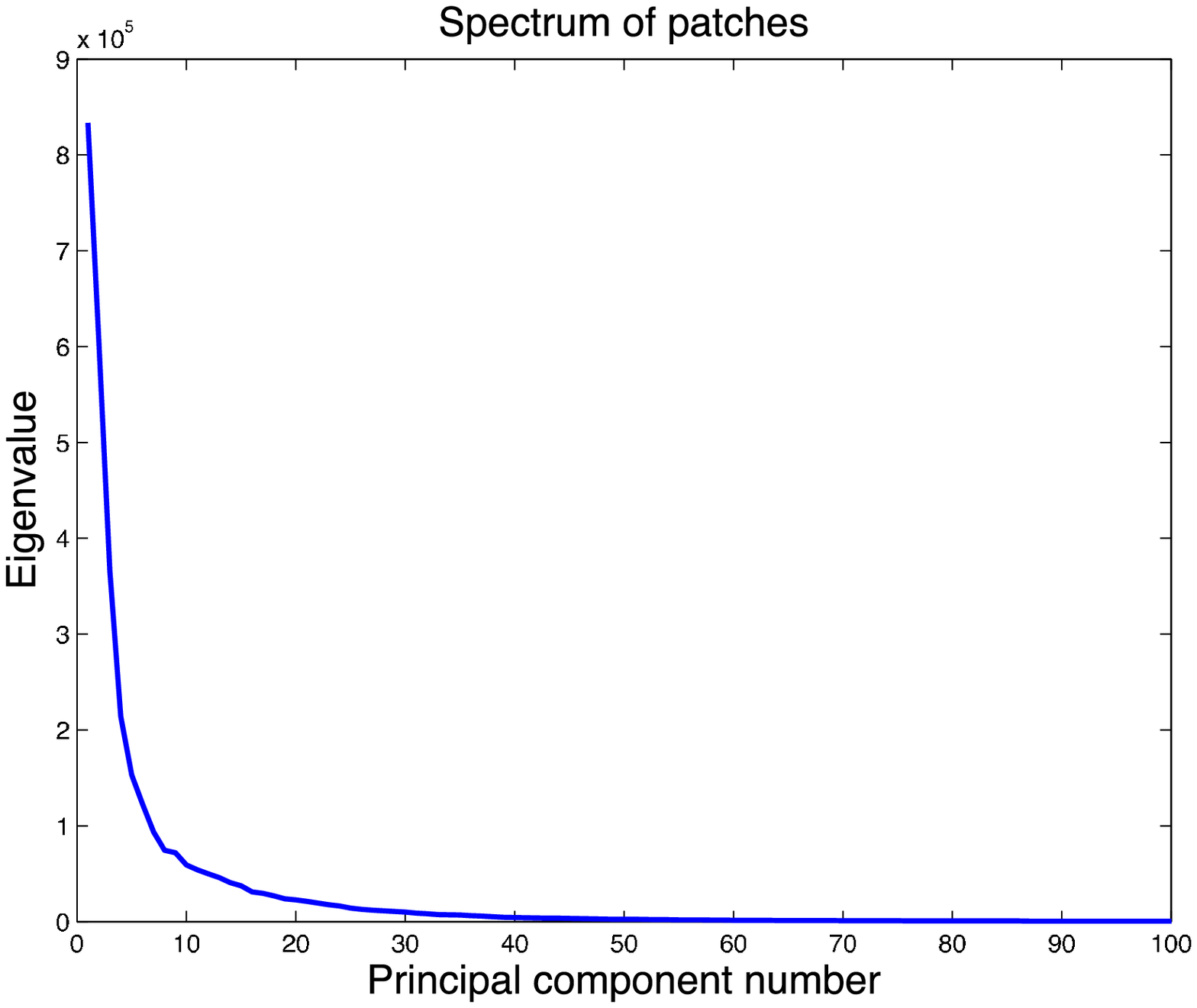}
}
\end{center}
\vspace{-7mm}
\caption{{\em Left:} The top $K=20$ eigenvectors, each reshaped to be
  $\Lambda \times R$ (after taking the middle $\phi$ slice to ease
  visualization).  {\em Right:} Associated eigenvalue spectrum of the
  data. Note that 50 or so eigenvectors are sufficient to capture most
  of the energy of the signal, showing that the speckle pattern, while
  complex, is inherently low-dimensional in the polar patch space.}
\label{fig:spectrum}
\end{figure}

To build the PCA model, we reformat the training set of patches into a
2D data matrix of size $n_{samples} = ((\Theta-2\delta \theta -1)
\cdot N)$ samples by $n_{dims} = (R \cdot \Phi \cdot \Lambda)$ dimensions. We
then center the data (subtract off the mean over all samples) and
build the covariance matrix (of size $n_{samples} \times
n_{samples}$). We then compute the eigenvectors of the
covariance matrix using singular value decomposition (SVD), which returns an eigenvalue-ranked
list of eigenvectors.  Each eigenvector can be reformatted back into a
$[R \times \Phi \times\Lambda]$ synthetic patch, as shown in
\fig{spectrum}(left). \fig{spectrum}(right) shows the sorted
eigenvalues, revealing that most of the variance is captured with $50$
or so components. Using this representation, the model for the
patches is arbitrary linear combinations of the first $K$
eigenvectors, where $K$ is a control parameter that dictates the freedom of
the model. %DISCUSSION ABOUT COMPANIONS IN TRAINING SET.

%  As per usual, the PCA step makes no use of any uncertainty
% estimates for the input data; the first $K$ eigenvectors span the
% $K$-dimensional subspace of the full data space that contains most of
% the variance of the original data. \fig{spectrum} shows the
% eigenvectors (reshaped into synthetic patches), as well as the sorted
% eigenvalues for a typical training set. 

Given the $K$ top eigenvectors from the training set, each patch in
the test set can be reconstructed by finding the linear combination
that minimizes the total squared reconstruction error (residual). The
optimal patch reconstruction $\hat{p}^{d,\theta,n}$ is given by
$(VV^T(p^{d,\theta,n}-\mu^{d,\theta}))+\mu^{d,\theta}$, where $V$ is a matrix of eigenvectors (of size
$[(R \cdot \Phi \cdot \Lambda) \times K]$), $\mu^{d,\theta}$ is the
mean over all training patches and $p^{d,\theta,n}$ is a
vectorized test patch. Reshaping the reconstruction back into a 
$[R \times \Phi \times\Lambda]$ patch, we compute the residual error
$\Delta P^{d,\theta,n} = (\hat{P}^{d,\theta,n}  - P^{d,\theta,n})$.

% Again, at this reconstruction set, we do not use any uncertainty or
% noise estimates for the data.  We perform this reconstruction for each
% patch $P_{r \theta n}$ at a number of different values of $K$, saving
% the patch residual $\Delta P^{(K)}_{r \theta n}$, which has the same
% dimensions as the original patch but contains the residual data minus
% reconstruction.

\begin{figure}[h!]
\begin{center}
\includegraphics[width=6.5in]{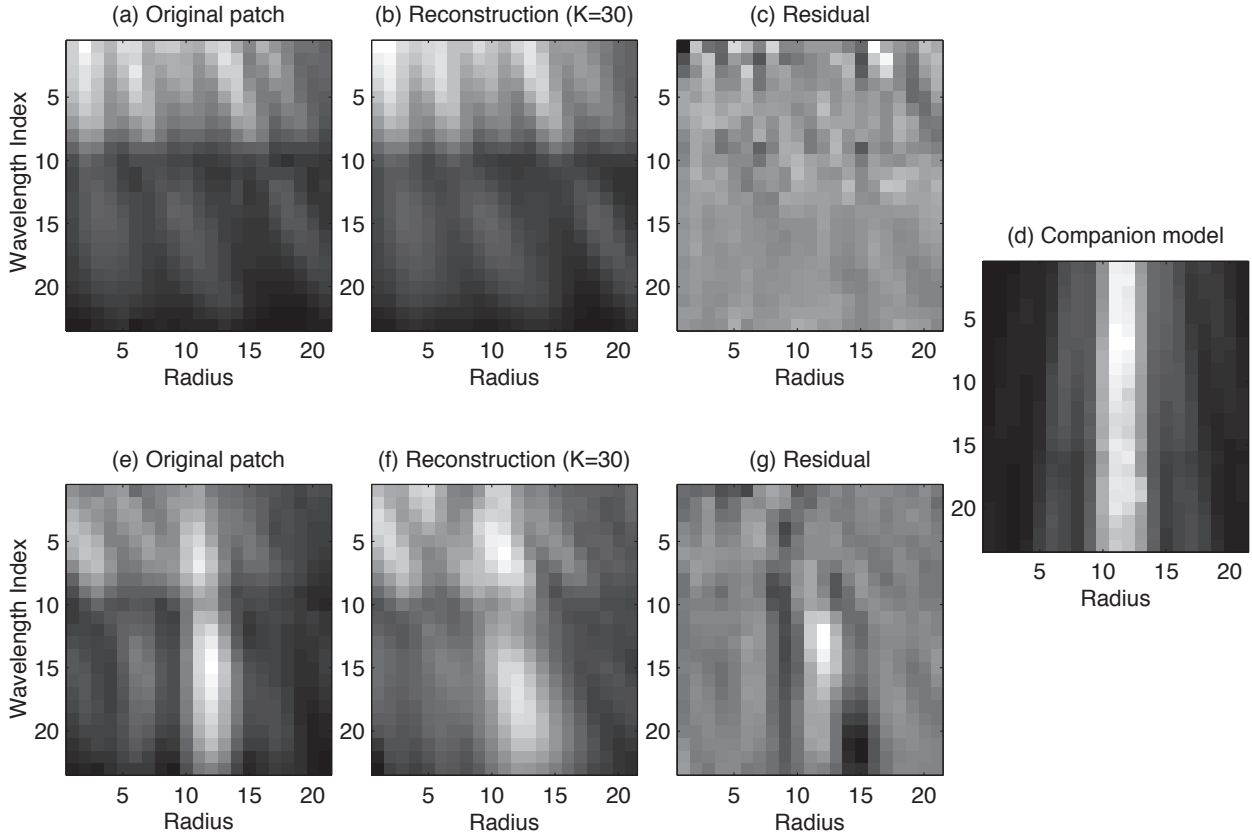}
\end{center}
\vspace{-7mm}
\caption{Reconstructions using our PCA model with $K=30$ components (averaged over angle $\phi$). {\em (a):} Typical
  patch, containing only speckles. {\em (b):} Reconstruction
  using our PCA model with $K=30$. {\em (c):} Error
  residual. Note the lack of structure. {\em (d):} Companion model,
  which has low correlation with (c). Note the Airy rings around the
  core, that spread
  out with wavelength. {\em (e):}  Patch
  containing companion. {\em (f):} Reconstruction from our
  model. {\em (g):} The error residual shows clear structure
  associate with the companion, i.e.~the PCA speckle model cannot
  reconstruct well the companion signal. This has high correlation
  with the companion model of (d).}
\label{fig:patch_recon}
\end{figure}

If there happens to be a companion centered at location $(d, \theta,
n)$ and if the model is not so free (for at least some values of $K$)
that it can fully absorb it, then the residual $\Delta P^{d,\theta,n}$ ought to look something
like a companion, since the vertical structure of the companion cannot
be modeled well by eigenvectors that consist of mainly diagonal
structure. If there is no companion, then the residual should be small
in magnitude and have little structure. These two scenarios are illustrated in
\fig{patch_recon}. 

The rationale behind splitting the data into separate train and test
sets is that if the same data was used for both roles, then the model
would easily over-fit the data, leaving minimal residual even if a
companion was present. Holding out patches around the test location
when building the PCA basis ensures that the test data is unseen, thus
the model is forced to generalize from the training data when
reconstructing the test patches, rather than memorizing its
peculiarities. The importance of this hold-out procedure is
demonstrated in \fig{heldout}. However, as we exhaustively search all $d,\theta$ locations, the
training and test sets are different for each and thus the PCA
basis must be recomputed each time. In practice, this is the
computational bottleneck of the scheme, since SVD takes some seconds to
compute the eigenvectors given that $n_{dim} ~ 2000--3000$.

\subsection{Correlation with Companion Model}

As illustrated in \fig{patch_recon}, the PCA model should leave some
structure of the companion (if present) in the residual signal. We
therefore cross-correlate the residual patches $\Delta
P^{d,\theta,n}$ with a \emph{companion model} $C$. 
This is a patch of dimension $[R \times \Phi \times
\Lambda]$ containing a centered point-spread function, obtained by
polar transforming the cartesian calibration data (using the
appropriate Jacobian). Since the spectrum of the companion is not known
ahead of time, a ``white spectrum'' is used for the companion model,
i.e.~uniform intensity at each wavelength band. The companion model is
shown in \fig{patch_recon}(d). 
 
\begin{figure}[h!]
\begin{center}
\mbox{
\includegraphics[height=3in]{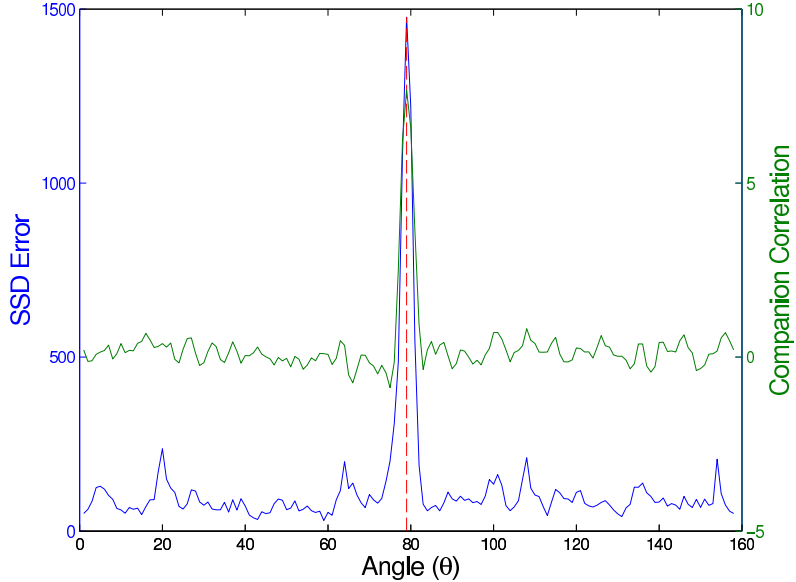}
}
\end{center}
\vspace{-7mm}
\caption{A detection signal (green) $o(d,\theta)$ for the
  example annular region shown in \fig{mean} at radius $d$. The
  sum of the squared residual error is shown in blue. Both curves have
a distinct peak at the angle corresponding to the location of the
companion (red).}
\label{fig:heldout}
\end{figure}

The amplitude of the cross-correlation signal is averaged over cubes
$n$ to give a detection signal at location $d, \theta$. In practice, we
find that normalizing the residual patches to have unit $\ell_2$
length performs better since it avoids a bias towards smaller values
of $d$ where the residuals tend to be larger. The normalized detection
response $o(d,\theta)$ is computed as:
\begin{equation}
o(d,\theta) = \sum_{n=1}^N \frac{\left(
  \sum_{r=1}^R \sum_{\phi=1}^\Phi  \sum_{\lambda=1}^{\Lambda} \Delta P^{d,\theta,n}_{r,\phi,\lambda}
  \cdot C_{r,\phi,\lambda}  \right)}{\|\Delta
P^{d,\theta,n}_{r,\phi,\lambda} \|_2 }
\label{eqn:corr}
\end{equation}
\fig{heldout} shows the detection response $o(d,\theta)$ and residual
error magnitude for the example region shown in \fig{mean}.
When computing $o(d,\theta)$, it is also possible to average over the
$\Phi$ patches that overlap with a given angle $\theta$. However, we
found that this tended to blur out the detection signal.

Finally, the detection response $o(d,\theta)$ exists in the polar domain and we
transform it back the cartesian domain to produce a final detection
map $m(x,y)$.

\subsection{Implementation}

Pseudo-code for the algorithm is given in Algorithm 1. A Matlab
implementation of the algorithm is provided on the project webpage:
\url{https://p1640.amnh.org/p1640/software/S4/}, with
accompanying instructions and examples. Beyond the core Matlab
package, it also requires the Image Processing toolbox and runs on
Windows, Mac and Linux platforms. The code is also compatible with
Octave, an open-source version of Matlab. The run-time of the
algorithm is a around 8 hours for a given star using typical settings, on a
fast PC and has a modest memory requirement (a few Gb).

%%%%%%%%%%%%%%%%%%%%
% This has already been updated!!!!!!!!
\begin{algorithm}[t!]
\small
\begin{algorithmic}[1]
\REQUIRE Data volume $I$ of dimension $[X \times Y \times \Lambda \times N]$, Companion model $C$
\REQUIRE \# Components $K$  
\REQUIRE Patch width $\Phi$, Ring width $R$, Test region half-width $\delta \theta$ 
\STATE Spatially center $I$
\FORC{$d=1$ to $(X/2-R)$}{ \%\% Loop over radius}
\STATE $J=$ Polar transform of annular region of $I$ at radius $d$, width $R$  
\STATE $P=$ Patchify $J$ using patches of angular width $\Phi$
\FORC{$\theta=1$ to $\Theta$}{ \%\% Loop over angle}
\STATE $P_{test} = P$ in region $\pm \delta \theta$ around $\theta$
\STATE $P_{train} = $ all $P$ not in $P_{test}$
\STATE Compute largest $K$ eigenvectors $V$ and mean $\mu$ from $P_{train}$
\STATE $\hat{P}_{test} =$ reconstruction of $P_{test}$ using $V$ and $\mu$
\STATE Compute residual $\Delta P= \hat{P}_{test}(\theta) - P_{test}(\theta)$
\STATE Normalize $\Delta P$ to unit length
\STATE $o(d,\theta) = $ correlation of $\Delta P$ with companion model
$C$ (\eqn{corr})
\STATE $m(x,y) = $ inverse polar transform of $o(d,\theta)$ 
\ENDFOR
\ENDFOR
\STATE Output: Detection map $m$
\end{algorithmic}
\small
\caption{The {\em S4} Algorithm}
\small
\label{alg:am} 
\end{algorithm}

\section{Results on Synthetic Data}
We initially demonstrate our approach using companions inserted into
real several stars recorded from the P1640
instrument: Alcor, HD87696 and GJ758. The input to our algorithm
consists of data cubes from the extraction and spectral calibration
pipeline, numbering $5$, $10$ and $15$ for the three stars,
respectively. 

Our evaluation consists of inserting fake companions of varying
intensity into data cubes and measuring the statistical significance
of peak at the true location. GJ758 and HD87696 have no detectable genuine
sources\footnote{GJ758 has a stellar companion (see
  \cite{Thalmann10}), but this is outside the field-of-view for our
  data.} that might
confuse the evaluation. Alcor does have a companion (Alcor B) but it
is sufficiently far from the star that it has been cropped from the
data cubes we use for testing our algorithm. We use a fake companion
that has a very similar spatial structure to the companion model $C$,
but differs in spectrum: the fake has a realistic reddish spectrum
while model $C$ is white.

We quantify the strength of the inserted source by measuring its mean
brightness over a 5 by 5 spatial window {\em relative} to the mean speckle
brightness in 5 by 5 window at a given location, averaging over
wavelength and cube. Put another way, under this measure 
a 5\% inserted companion at two different locations has different
absolute brightness, but the same companion/speckle brightness ratio.

We characterize the statistical significance of the detected peak in
two ways. First, we measure how many standard deviations it is above
the rest of the detection map. The background of the detection map has
statistics that are close to Gaussian, making this a valid
comparison. Second, we show the rank of the peak, relative to all
other elements in the map. A rank of 1 indicates that it is the
strongest peak in the entire map. This test does not make any
assumptions about Gaussianity of the data and is similar in spirit to
other non-parametric tests such as the Wilcoxon-rank sum test.

All experiments used the following settings: $R=30$ pixels, $\delta
\theta=10$ pixels and $\Phi=3$ pixels. Performance is relatively
insensitive to the first two of these. $\Phi=3$ is superior
to $\Phi=1$ and similar to $\Phi=5$. But $\Phi=5$ is not preferred as
it significantly increases the dimensionality of the PCA space, which
slows the algorithm considerably.  

\begin{figure}[h!]
\begin{center}
\includegraphics[width=6.5in]{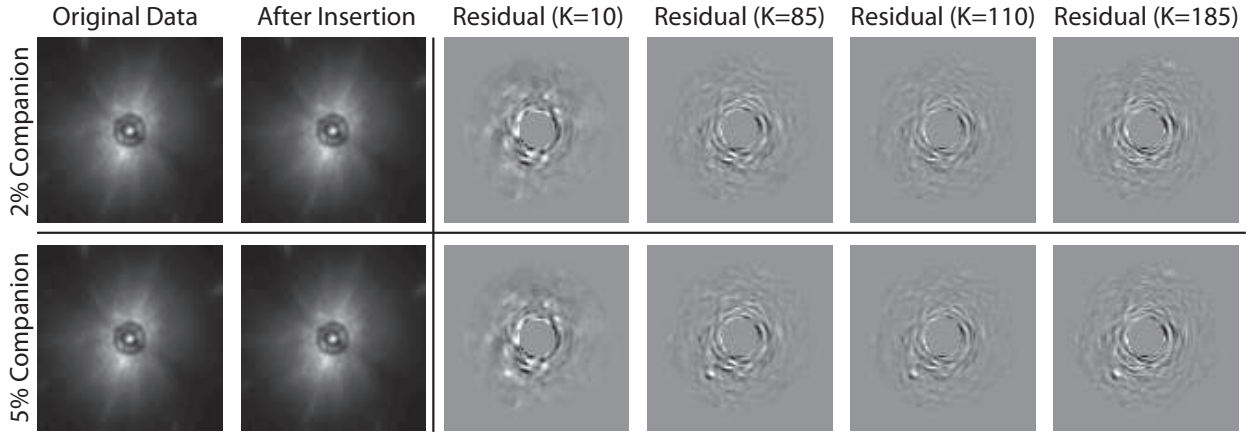}
\end{center}
\vspace{-7mm}
\caption{A single cube of HD87696 (averaged over wavelength), along with PCA
  residuals for fake companions of $2\%$ and $5\%$ relative intensity
  (top and bottom rows, respectively), inserted at the 7 o'clock
  position. Columns from L to R: (i)
  Original cube. (ii) Cube after insertion of fake
  companion. (iii)-(vii): PCA residuals $\Delta P$ mapped back to a
  cartesian representation for $K=\{10,85,110,185\}$. The dynamic
  range used in plotting decreases across the row (but is constant
  within a column), since increasing $K$ reduces the residual magnitude. Note: (a)
  invisibility of companion after insertion (compare columns (i) \& (ii)). (b) $5\%$
  insertion is visible in residuals, while $2\%$ is not. Correlating
  these residuals with a companion model makes detection below $2\%$  possible (see
  \fig{maps_hd87696} and \fig{sens_map}). (c) Too few components
  ($K=10$) or too many ($K=185$) reduce the contrast of the
  companion relative to speckle residuals, thus impairing detection performance. 
}
\label{fig:residuals}
\end{figure}

The number of components in the PCA model, $K$, is an important
parameter in the algorithm and for each experiment we vary this
systematically and show detection maps $m$ for the best performing
value, which is typically around $K=100$. If $K$ is too small, the PCA
model will not be able to reconstruct the speckles well. If $K$ is too
large it will also reconstruct the companion, leaving nothing in the
residual. See \fig{residuals} for an illustration of this.
\fig{residuals} also demonstrates how the fake sources are inserted
and what the PCA residuals look like, before correlation with the
companion model. 

\begin{figure}[h!]
\begin{center}
\includegraphics[width=6.5in]{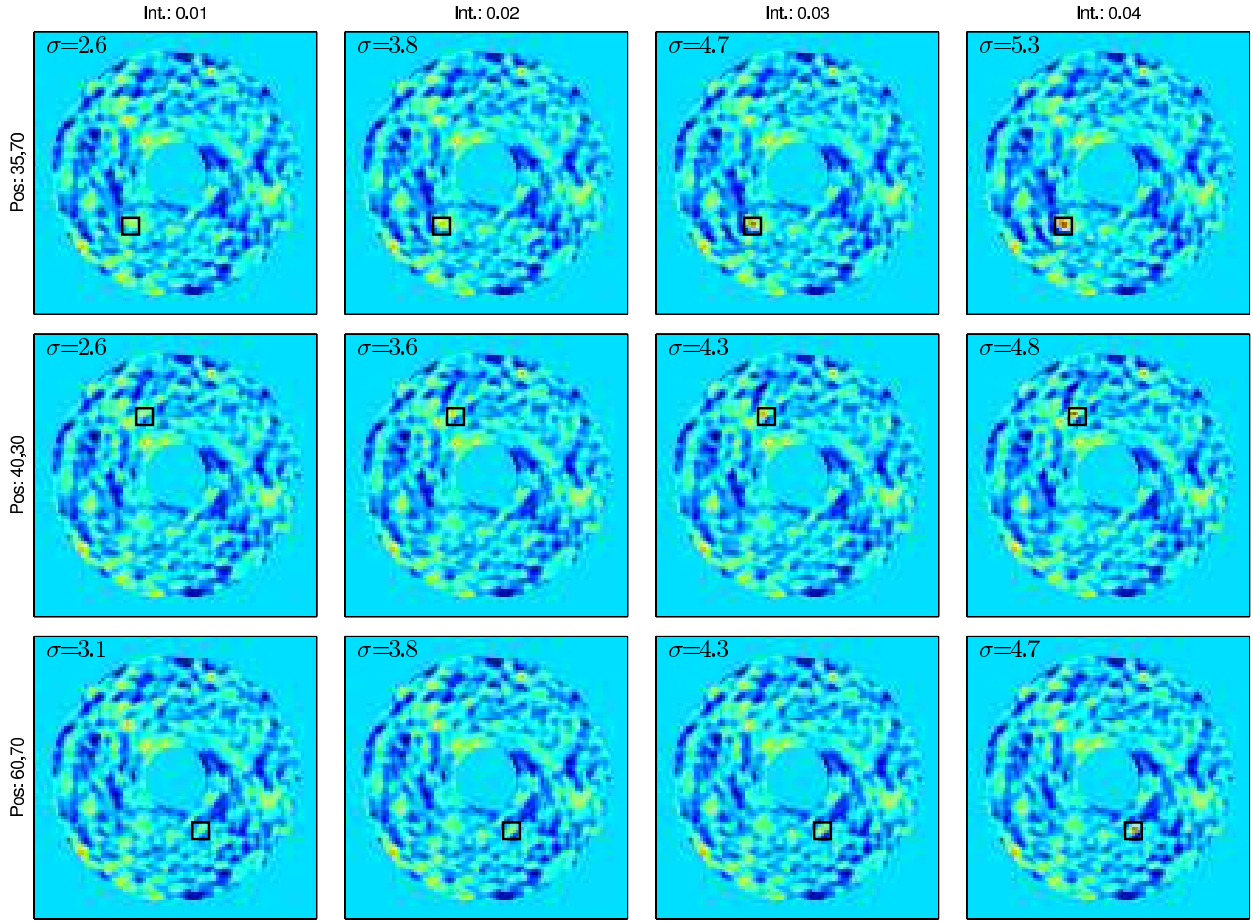}
\end{center}
\vspace{-7mm}
\caption{Cartesian detection output maps $m$ for the star HD87696, for fake
 companions inserted at 3 different locations at 1,2,3 and 4\%
 relative intensities. Red and yellow corresponds to a high response, blue
 corresponds to a low response. The cyan areas are not examined due
 to boundary issues caused by the width of the annular section.
 The black rectangle shows the true location of
the inserted source. Each plot gives the statistical significance of
the detection. The algorithm is able to reliably detect the
companion down to between 1-2\% of the speckle flux. The figure is
best viewed in electronic form. }
\label{fig:maps_hd87696}
\end{figure}

\fig{maps_hd87696} shows the fake sources added to $N=5$ cubes of star
HD87696. Both the location and brightness of the source are varied,
the latter from 1--4\% of the speckle brightness at the insertion
location, thus being a relative, not absolute, measure of detection
performance. \tab{hd87696} presents these results in a different form,
showing the number of standard deviations above the background, as
well as the rank of the true companion location versus other local
maxima in the response map. Taking a 3$\sigma$ threshold, we see that
the algorithm can reliably detect the companion down to
around 1-2\% of the speckle brightness. 

\begin{table}
\small
\begin{center}
\begin{tabular} { | c | c | c || c | c | c | c | c | c | c | c | } \hline
 Location & X & Y & 0.5\% &  1\% &  2\%  &  3\%  & 4 \% & 5\% & 7.5\%
 & 10\% \\ \hline \hline
 1    &   35  &    70  &  0.7 / 32  &  1.9/  19  &   3.9/ 1  &  5.4/   1  &  6.6/   1  &
 7.4/1   &    8.7/1  &    9.3/ 1\\ \hline
2    &   40  &    30  &  2.1/7  &   2.8/ 7  &  4.1/  1  &   5.2/  1  &  6.1/   1  &    6.9/ 1
 &   8.3/ 1  &   9.2/  1\\ \hline
3    &   60  &    70  &   -  &    2.7/ 6  &   4.4/ 1  &  5.7/   1  &   6.8/  1  &    7.6/ 1
 &    8.7/1  &   9.1/  1\\ \hline
 4    &   65  &   35  &   1.2/28  &   1.9/22  &   3.1/ 3  &    4.1/ 1  &    5.0/ 1  &
5.7/1   &   6.8/ 1  &   7.4/  1\\ \hline
\end{tabular}
\end{center}
\caption{Synthetic companion insertion on HD87696. For
 4 different locations and fake source intensities ranging from
 $0.5\%$ to $10\%$, relative to the local speckle level. For each
 combination, we give the number of standard deviations above the
 background, as well as the rank of the peak over the entire map. A
 rank of 1 means that the strongest local
 maxima in the detection map is at the true location of the inserted
 source (i.e.~it has correctly found the companion). Rank $n$ means
 that the local maxima containing the companion is the $n$ strongest in
 the detection map. Our algorithm is reliable
 (i.e.~$\geq 3\sigma$) for companion brighter than 2\% of the speckle
 flux.}
\label{tab:hd87696}
\end{table}

\begin{figure}[h!]
\begin{center}
\includegraphics[width=6.5in]{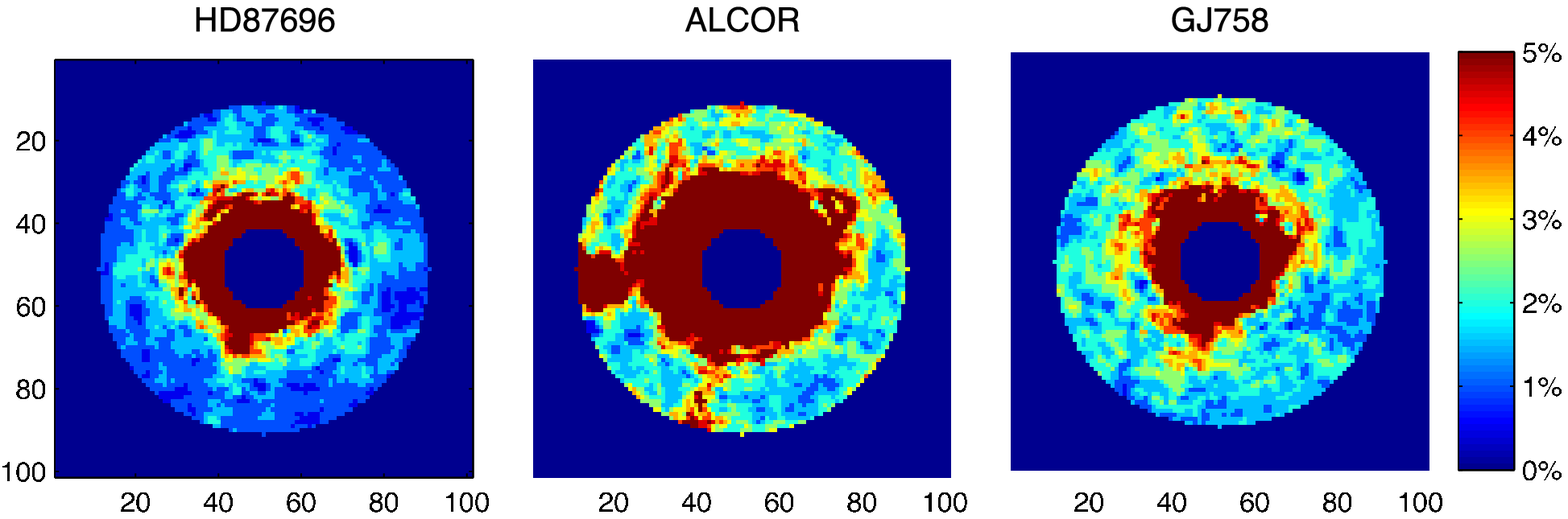}
\end{center}
\vspace{-7mm}
\caption{Cartesian detection sensitivity maps for stars HD87696, Alcor
  and GJ758. The color shows the detection threshold at each
  location, as a fraction of the speckle flux at that location. Red
  corresponds to a high response, blue corresponds to a low
  response. Dark blue regions are those non-examinable due to the
  width $R$ of the annular region. The performance of the algorithm
  depends on the quality of the cubes input to the algorithm, which
  was noticeably worse for Alcor than the other two stars. In the
  case of HD87696, which has the best quality cubes, large portions of
the map are around (or below) 1\%.}
\label{fig:sens_map}
\end{figure}

By assuming that no genuine companions are present in the detection
maps $m$, any strong peaks in the detection signal must be due to
noise. This allows us to compute a sensitivity map which records at
each location the brightness level of a companion whose detection
signal in $m$ just exceeds the largest magnitude noise
peak. \fig{sens_map} shows sensitivity maps for three different stars,
HD87696, Alcor and GJ758. The maps show the sensitivity is mostly at
the the 1-2\% level, consistent with the other results. In areas near
the occulting disk, however, the sensitivity is poor (worse than
10\%). The performance of the algorithm is highly dependent on the
quality of the input data, which varied between the stars.

\begin{figure}[h!]
\begin{center}
\includegraphics[width=3.2in]{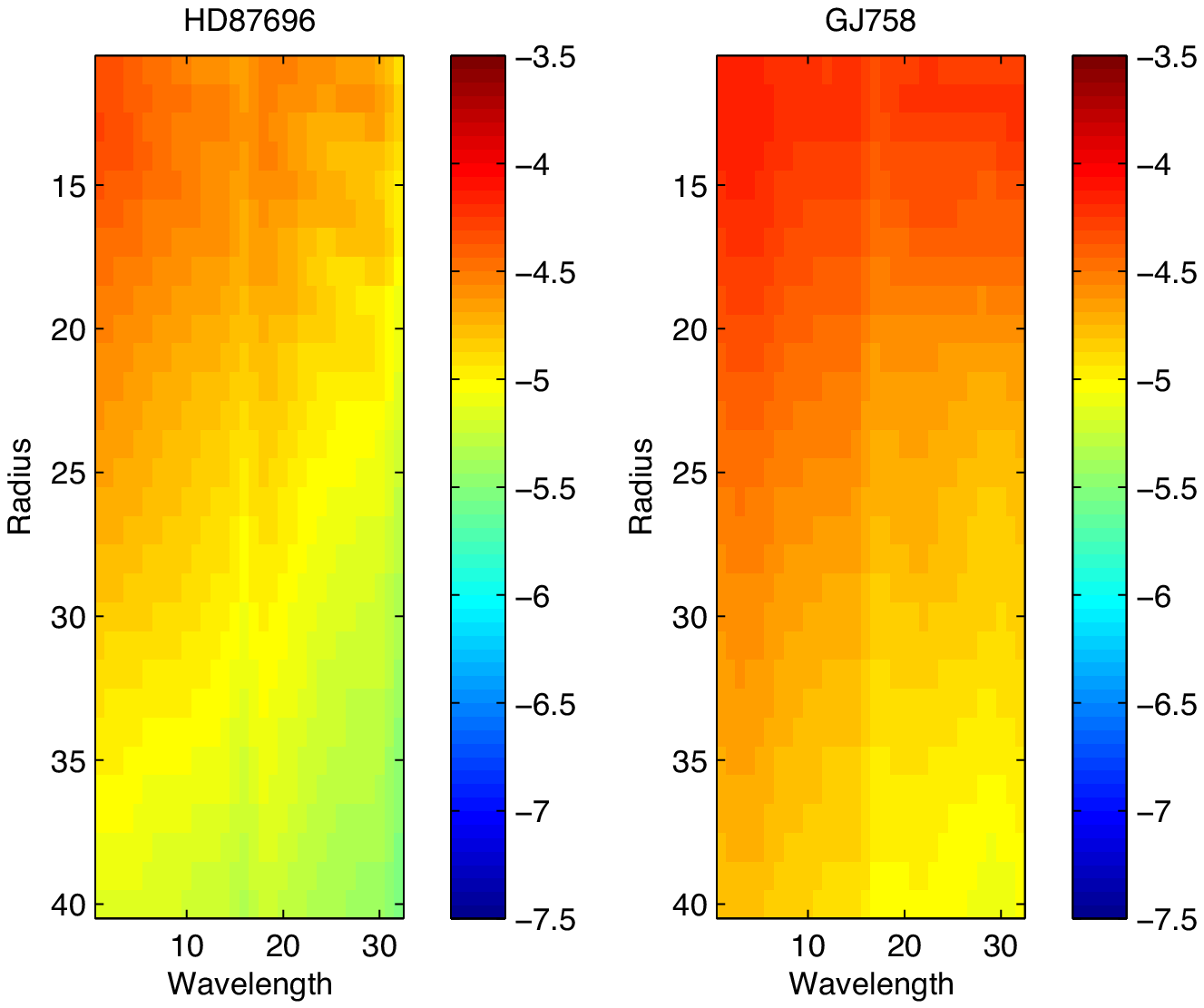}
\includegraphics[width=3.2in]{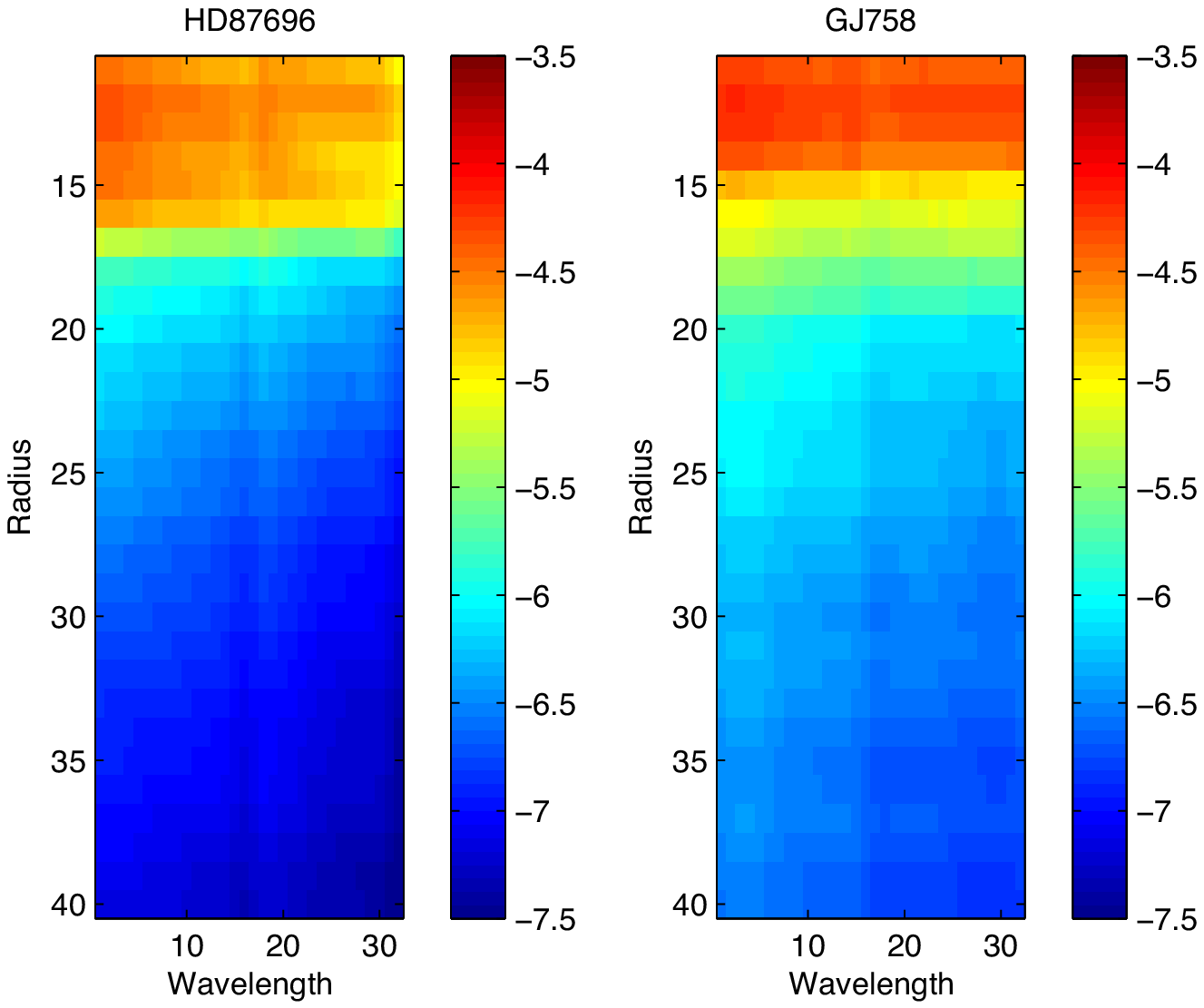}
\end{center}
\vspace{-7mm}
\caption{Absolute contrast as a function of radius and wavelength
  before (L) and after (R) the application of S4 to two stars, HD87696
  and GJ758. Shown in $\log_{10}$ scale. S4 give around 2 orders of
  magnitude improvement, reaching  $10^{6-7}$ absolute contrast for outer
radii. }
\label{fig:abs_contrast}
\end{figure}

The sensitivity maps in \fig{sens_map}, allow us to derive a
measurement of absolute contrast. At a given location this is done by
combining the sensitivity (i.e.~minimum detectable companion
brightness, relative to the background speckles) with the ratio of the
speckle brightness to the unocculted star. Averaging over each annulus
gives the plots shown in \fig{abs_contrast}, which show the absolute
contrast as a function of radius and wavelength before and after the
application of the S4 algorithm to the P1640 data from HD87696 and GJ758.
S4 can be seen to give an improvement in detection sensitivity of around 2 orders
of magnitude, reaching $10^{6-7}$ contrast for outer
radii. Furthermore, S4 compensates for the chromatic behavior of the
coronagraph, as shown by the softening of the contrast's dependence
on wavelength.

\begin{figure}[h!]
\begin{center}
\includegraphics[width=6.5in]{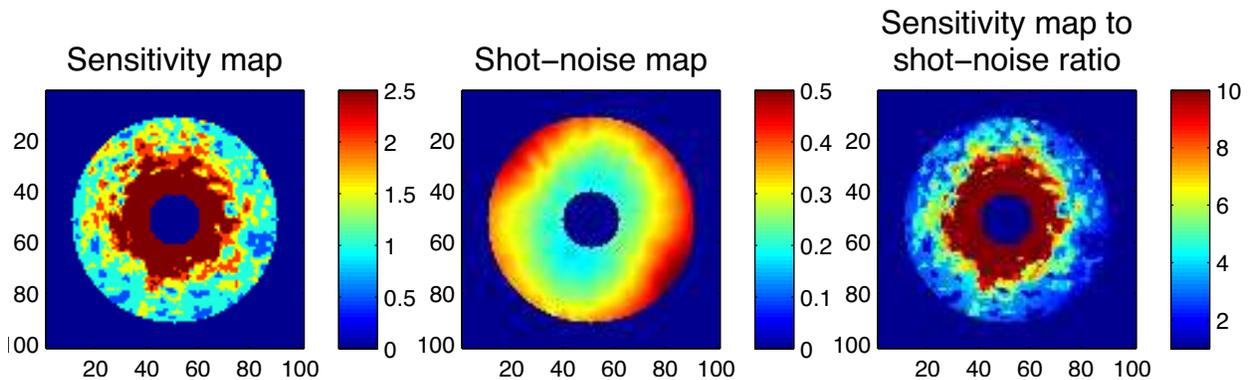}
\end{center}
\vspace{-7mm}
\caption{The detection capability of our algorithm, compared to the physical
  limit imposed by shot-noise in the detector, for star HD87696. {\em
    Left:} Detection sensitivity map (same as \fig{sens_map}(left),
  but different dynamic range). Units are $\%$ relative to the speckle
  intensity. {\em Middle:} Shot noise map, computed using
  \eqn{shot}. Units are $\%$ relative to the speckle
  intensity. {\em Right:} Ratio of the sensitivity map to the
  shot-noise limit. In places, our algorithm is close to the
  shot-noise limit (i.e~ratio $<2$), but typically it is within a
  factor of $4$, apart from regions close to the star.}
\label{fig:shot}
\end{figure}

Shot-noise due to photon arrival times and the total exposure time
provides a physical lower-limit to companion detection. We compute the
absolute 2D shot noise map $s(x,y)$ as:
\begin{equation}
s = \sqrt{ T \cdot (\tilde{C} * \frac{1}{N} \sum_n \sum_\lambda
  I_{x,y,\lambda,n} )}  
\label{eqn:shot}
\end{equation}
where $T$ is the exposure time (secs), $*$ is a 2D correlation
operation and $\tilde{S}$ is a 2D binary mask of the companion model,
where all but the brightest central core of the wavelength-averaged
companion model $C$ has been thresholded to zero. The correlation has
the effect of summing over all pixels used in the companion model (the
companion occupies more than a single pixel). In \fig{shot}, we plot
the ratio of $s$ to the data $I$, to give the same relative measure
used when producing the sensitivity maps shown in \fig{sens_map}. In
the case of HD87696, we see that the shot-noise limit is $\sim 0.4\%$
around the periphery and $0.25-0.3\%$ near the star. Our algorithm's
sensitivity map of HD87696 has large regions within a factor of $4$ of
the shot-noise limit. However, close in to the star, the our
sensitivity drops to $>5\%$, considerably worse than the limit.

\subsection{Comparison to LOCI}

\begin{figure}[h!]
\begin{center}
\includegraphics[width=6.5in]{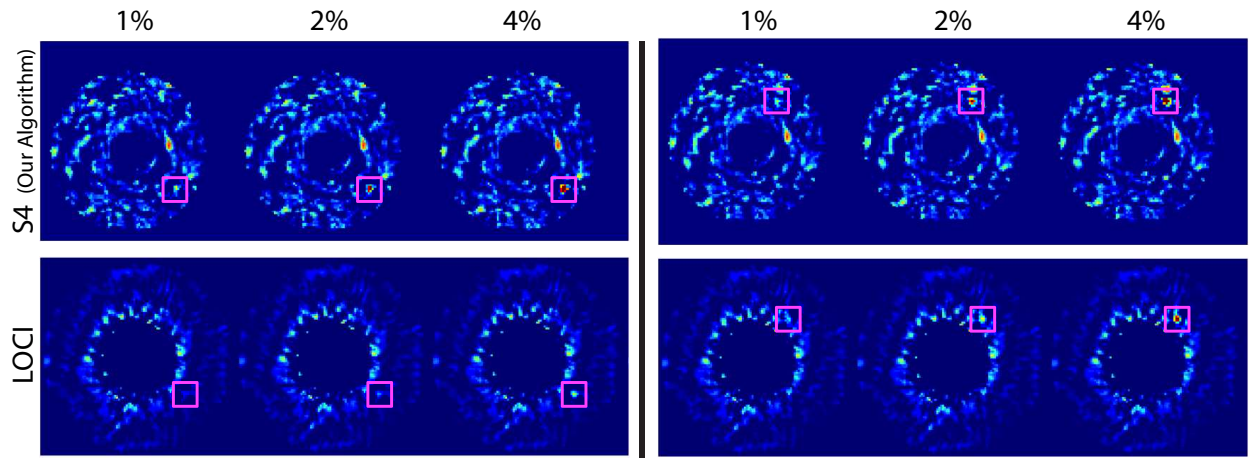}
\end{center}
\vspace{-7mm}
\caption{Comparison to LOCI (\cite{Pueyo11}). Fake companion insertions (3
  intensities) at two different locations. Red = higher detection
  level. The true location is indicated by the magenta box. Our
  algorithm ({\em S4}) cleanly locates the companion at the 2\% level, which is
  still visible at the 1\% level. LOCI finds the 4\% companion but the
  2\% one is no longer brighter than background artifacts.}
\label{fig:loci}
\end{figure}

We also perform a direct comparison to the damped LOCI algorithm
\cite{Pueyo11}, an improved variant of the original algorithm
(\cite{Lafreniere07}). 
Data cubes with fake companion insertions at
five randomly chosen locations were created and provided to the
authors of \cite{Pueyo11}. They ran
their damped LOCI implementation on the data cubes and returned
the results, enabling their evaluation under the metrics used
above. The cubes were also presented to our algorithm, using the
settings described above. Thus, both the input data and the evaluation
metrics were the same, the only difference being the detection
algorithm itself. \tab{loci} shows compares the companion rank of the
two algorithms for different brightnesses of fake companion. At the
2\% level, {\em S4} has significance values $\geq 3.9\sigma$ for all
five companion locations, where are LOCI drops to $1.8\sigma$ for some
positions. At the 1\% level, neither algorithm reliably finds the true
location, but {\em S4} has a mean significance (averaging over the 5
locations) of $2.5\sigma$, compared to $1.8\sigma$ for LOCI. A visual
comparison is shown in \fig{loci}.

\begin{table}[h!]
\small
\begin{center}
\begin{tabular} { | c | c |c| c || c | c | c | c | c | c | c | c | } \hline
 Location & X & Y & Method & 1\% &  2\% &  4\%  &  6\%  \\ \hline \hline
 1    &   26  &    72 & {\em S4}: &  2.9 / 3  &   5.2 / 1  &  7.2 /  1&  7.8 /   1  \\ \hline
      &     &              & LOCI: &   1.8 / 31  &  3.0 /  10  &  5.7 /  1  & 8.3/    1   \\ \hline  \hline
2    &  30  &   40 & {\em S4}: &   4.2 / 2  &  5.5 /   1  &  6.9 /  1
&  7.4 /  1   \\ \hline
      &     &              & LOCI: &   3.1 / 2  &  4.6 /  1  & 7.4/
      1  &  10.1 /   1 \\ \hline  \hline
3    &  56  &   72   & {\em S4}: &  2.2 / 3  &  3.9/  1  &   6.0/ 1  &   7.0/  1  \\ \hline
      &     &              & LOCI: &   0.8  / 7  &    -  &   7.5/ 1  &     -  \\ \hline  \hline
4    &  67  &   30   & {\em S4}: &  2.3 /  6  & 4.5/  1  &  7.0 /  1
&  8.0 /   1  \\ \hline
      &     &              & LOCI: &   2.2/ 23  &  3.8 /  3  &  6.9/  1  &     -    \\ \hline  \hline
5    &  75  &   70   & {\em S4}: &   1.1/ 5  &  4.1/  1  &  7.9 /  1  &  8.5/   1   \\ \hline
      &     &              & LOCI: &   1.1 / 56  &   1.8 /  31  &  3.5
      /16  &   5.1/  1    \\ \hline  \hline
\end{tabular}
\end{center}
\caption{A comparison between damped LOCI (\cite{Pueyo11}) and our algorithm
  ({\em S4}) for five different locations on star HD87696. We give the
  number of standard deviations above the background (1st number)
  and rank (2nd number) of the true peak. {\em S4} typically has
  higher significance values (and lower rank) than LOCI.}
\label{tab:loci}
\end{table}

\section{Results on HR8799}

We applied our algorithm to 10 cubes of data acquired over two consecutive
nights (14/15th June 2012) of the star HR8799. Observations with the
Keck telescope by Marois et al. have revealed 4 companions orbiting
the star, making it an ideal test case for our algorithm.

\begin{figure}[h!]
\begin{center}
\includegraphics[width=6.5in]{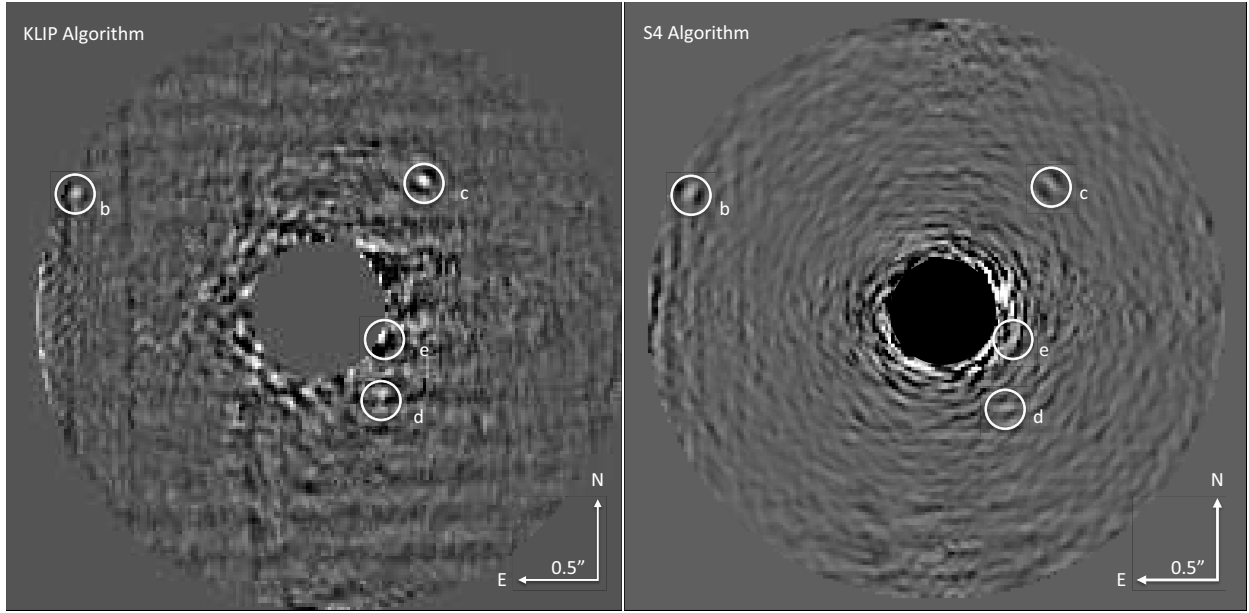}
\end{center}
\vspace{-7mm}
\caption{From \cite{Oppenheimer13}: HR8799 PCA residual maps for the \project{KLIP} algorithm
  \cite{Soummer12} (L) and S4 (R). While (b) and (c) companions are
  clearly visible in both maps, the (d) companion is clearer in
  S4. The (e) companions are weak in both. Note that for S4, this
  residual map is only an intermediary output. See \fig{hr8799out} for
the final output.}
\label{fig:hr8799}
\end{figure}

\begin{figure}[h!]
\begin{center}
\includegraphics[width=3.2in]{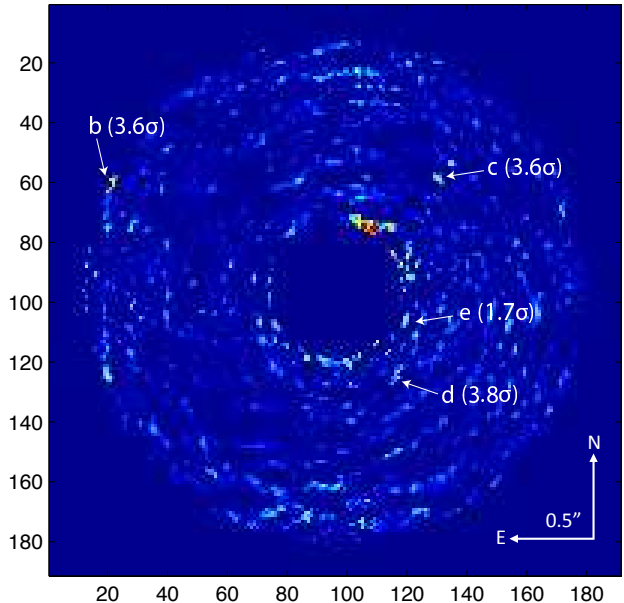}
\end{center}
\vspace{-7mm}
\caption{From \cite{Oppenheimer13}: The final output of S4 for HR8799. The companions have
  distinct peak, which for (b),(c) and (d) companions are
  significantly above the surrounding noise ($\geq 3.6 \sigma$).  Note
  also that the
  position of the companions is more precisely localized than in
  \fig{hr8799}(right).}
\label{fig:hr8799out}
\end{figure}

After centering the cubes, we applied the S4 algorithm using
parameters: $R=13$, $\Phi=3$ and $\delta \theta=10$. Pueyo \etal also
applied the KLIP algorithm \cite{Soummer12} to
the same data. The results of the two algorithms are shown in
\fig{hr8799} and \fig{hr8799out}, taken from \cite{Oppenheimer13}.
The former shows the PCA residual
map produced by both algorithms, with S4 giving higher signal to noise
than KLIP, enabling the clear detection of the (d) companion. Applying
the companion model to the residual produces the output map shown in
\fig{hr8799out}. This gives cleaner detections, with three of the peak
above $3\sigma$ significance. \cite{Oppenheimer13} also used S4 to
perform a spectral analysis of these companions and the algorithmic
details of this study will be described in a forthcoming paper.

% should be in final version
%SHOW DETECTION MAP WITH CORR, NORMCORR, SPECIAL.

%KLIP outperforms LOCI.

\section{Discussion}

We have introduced the {\em S4} algorithm for companion detection and
demonstrated it on several stars captured by the \project{P1640}
instrument. The approach outperforms the existing LOCI and KLIP
algorithms by a significant margin and approaches the shot-noise limit
for radii further from the star.  The performance of the algorithm is
dependent on the speckle structure being clearly visible within the
input data. Hence, in regions close to the star which lack clear
structure, the algorithm performance drops off significantly with a
relative sensitivity worse than $5\%$. When applied to the P1640 data,
S4 improves the absolute contrast by 2 orders of magnitude, to give
levels around $10^{6-7}$, beyond the core region.

There are several issues with the methods presented that leave room
for improvement: (a) We do not use a proper instrument noise model --
the PCA model could be modified to incorporate known noise properties
of the instrument (such as that of \cite{hmf}). (b) Our companion
model assumes a white spectral emission distribution (SED), which is
clearly wrong in general. A better approach would be to let this also
be unknown and fit both the speckle and companion SED simultaneously
to the data. (c) We only build our model from data cubes of a single
star. This severely limits the amount of training data available and
necessitates the use of simple models, such as PCA. A better approach
would be to draw statistical strength from observations of multiple
stars captured by the \project{P1640} instrument. While the speckle
patterns differ between stars, they undoubtedly contain similarities
that can be learned. For example, given enough data, the space of
likely atmospheric distortions could be learned. (d) Related to this,
the model we use does not take as input any meta data about the
telescope attitude (e.g.~ ``gravitational loading'' or ``differential
chromatic refraction'') or observing conditions in training.  A more
sophisticated model might leverage this information to learn
dependencies on these variables.  (e) If more data were available,
other modeling options become viable. For example, sparse coding
(\cite{Tibshirani96,Mairal2009}) is a more flexible model than PCA as
it can capture multiple low-dimensional linear subspaces (as opposed
to a single one). Another option would be discriminative approaches
based on support vector machines. Replacing PCA with these approaches
might deliver superior performance.  (f) The input data $I$ to our
algorithm is the result of a complex extraction pipeline whose details
we have not considered in this article. It is possible that this
pipeline incorporates a number of sub-optimal operations on the data
that inadvertently degrade the signal-to-noise. These might be avoided
if our model were to be directly applied to the raw sensor
measurements.

%NEW POINTS: OTHER PEOPLE USED ADI + LOCI. WE BEAT LOCI. 

Finally, while the algorithm has been designed for the \project{P1640}
instrument, it can easily be applied to data from other instruments
that operate on similar principles.  We are working on methods for
obtaining the true spectrum of the detected companions.  This will be
reported in a subsequent publication.

\section*{Acknowledgements}
RF \& DH are partially supported by NSF CDI
\#1124794. RF is also partially supported by a Sloan Fellowship.
Project 1640 has been funded by National Science Foundation grants AST-
0520822, AST-0804417, and AST-0908484.  The project also acknowledges the
generous support of Hillary and Ethel Lipsitz, the Plymouth Foundation and
Futdi.

\bibliographystyle{apj}
\bibliography{astro}

\end{document}